\documentclass[twocolumn,pra,aps,superscriptaddress]{revtex4-1}

\usepackage{lineno}
  \usepackage{mathptmx}
\usepackage{subfigure}
\usepackage{dcolumn}
\usepackage{amsmath,amssymb}
\usepackage{bm}
\usepackage{color}
\usepackage{overpic}
\usepackage{latexsym}
\usepackage{epstopdf}
\usepackage{color}
\usepackage[english]{babel}
\usepackage{latexsym}
\usepackage{stmaryrd}

\usepackage{psfrag,graphicx} 
\usepackage{epsf} 
\usepackage{subfigure} 
\usepackage{amsmath} 
\usepackage{amssymb} 
\usepackage{amsfonts}
\usepackage{bm}
\usepackage{natbib}
\usepackage{epstopdf}
\usepackage{appendix}

\definecolor{mygrey}{gray}{0.35}
\definecolor{myblue}{rgb}{0.2,0.2,0.8}
\definecolor{myzard}{cmyk}{0,0,0.05,0}
\definecolor{mywhite}{rgb}{1,1,1}
\definecolor{myred}{rgb}{1,0.,0.3}

\usepackage[colorlinks=true,citecolor=myblue,linkcolor=myred]{hyperref}
\def\be{\begin{equation}}
\def\ee{\end{equation}}
\def\ba{\begin{align}}
\def\enda{\end{align}}
\def\bi{\begin{itemize}}
\def\ei{\end{itemize}}

 \def\ee{\mathord{\rm e}}
 
 \def\ii{\mathord{\rm i}}

\def\half{\textstyle\frac{1}{2}}

 \def\ee{\mathord{\rm e}}
 
 \def\ii{\mathord{\rm i}}

\def\half{\textstyle\frac{1}{2}}

\renewcommand{\ii}{{\rm i}}
\renewcommand{\ee}{{\rm e}}

\def\beq{\begin{equation}}
\def\beq{\begin{equation}}
\def\eeq{\end{equation}}

 \newcommand{\ket}[1]{|#1\rangle}
 \newcommand{\bra}[1]{\langle #1|}

\begin{document}


\title[Short Title]{Dissipation-Assisted Quantum Information Processing with Trapped Ions }

\author{A. Bermudez}
\affiliation{Institut f\"ur Theoretische Physik, Albert-Einstein Alle 11, Universit\"at Ulm, 89069 Ulm, Germany}

\author{T. Schaetz}
\affiliation{Albert-Ludwigs-Universit\"at Freiburg, Physikalisches Institut, Hermann-Herder-Strasse 3, 79104 Freiburg, Germany}

\author{M. B. Plenio}
\affiliation{Institut f\"ur Theoretische Physik, Albert-Einstein Alle 11, Universit\"at Ulm, 89069 Ulm, Germany}

\begin{abstract}
We introduce a scheme to perform dissipation-assisted quantum information processing in  ion traps considering realistic decoherence rates, for example, due to motional heating. By means of continuous  sympathetic cooling, we overcome the trap heating by showing that the damped vibrational  excitations can still be exploited to mediate coherent interactions, as well as collective dissipative effects. We describe how to control their relative strength experimentally, allowing for protocols of coherent or dissipative generation of entanglement. This scheme can be scaled to larger ion  registers  for coherent or dissipative  many-body quantum simulations.

\end{abstract}

\maketitle

Among the  platforms  for quantum computation (QC) and simulations (QS)~\cite{qc_qs},  trapped ions~\cite{qc_qs_ions}  stand out as  excellent  small-scale prototypes, which have  a well-defined  roadmap towards  large-scale devices   based on micro-fabrication~\cite{microtraps, many_body_review}. The success  of this technology depends on the impact of various sources of decoherence, such as  the  anomalous heating induced by the electric  noise emanating from the trap electrodes~\cite{anomalous_heating}. The strong ion-ion couplings, required  for scalable   QC/QS, demand that the ions lie closer to the  electrodes of these miniaturized  traps,  where the heating is  critical and must be carefully considered. A strategy to minimize it is to cryogenically cool the  setup~\cite{electrode_cooling}, or to clean the electrodes  by  laser ablation~\cite{laser_cleaning} or ion bombardment~\cite{ion_bombardment_cleaning}.  Although these approaches are promising, a substantial residual noise still exists. We propose to minimize it by   applying sympathetic laser cooling  {\it continuously} during the whole QC/QS protocol. 

Sympathetic cooling requires active laser cooling of a  subset of ions, and passive cooling of the remaining ions   by  Coulomb interaction. This technique may  overcome the motional heating~\cite{sympathetic_kielpinski,gs_sympathetic_cooling_focusing}, and  has already been  implemented  between  sequential  gates for QC~\cite{sc_quantum_logic}. However, the larger heating rates of surface traps would require to  cool  also during the gates. There are different schemes along these lines: {\it (i)} In the absence of fluctuating electric gradients, interactions can be mediated by vibrational modes robust  to the heating, while continuously cooling the remaining modes~\cite{sympathetic_kielpinski}. {\it (ii)} By  using  far-detuned state-dependent  forces~\cite{ms}, the mediated interactions do not rely on the motional coherence, and can thus withstand a heating/cooling that is considerably weaker than the interactions. {\it (iii)} For  ground-state cooled crystals,  resolved-sideband cooling may  provide a dissipative force that improves the success/fidelity of protocols that are shorter than the inverse of the heating rate~\cite{Beige}. Unfortunately, these  requirements are not met in  current surface traps: {\it (i)} Since ions lie close to the electrodes,  electric gradients prevent the isolation of robust  modes. {\it (ii-iii)}  Heating rates in room-temperature setups ($1$ phonon/ms~\cite{phonon_hopping})   coincide with the above protocols speed~\cite{ms,Beige}. 

In this work, we propose a {\it dissipation-assisted protocol} based on an always-on sympathetic cooling that  overcomes the anomalous heating for surface traps. We identify regimes where the sympathetically-cooled vibrational modes can be  used as mediators of both coherent interactions and collective dissipation. Since we only require  Doppler cooling, this proposal  can be applied to larger  registers  for QC/QS.

\begin{figure}

\centering
\includegraphics[width=1\columnwidth]{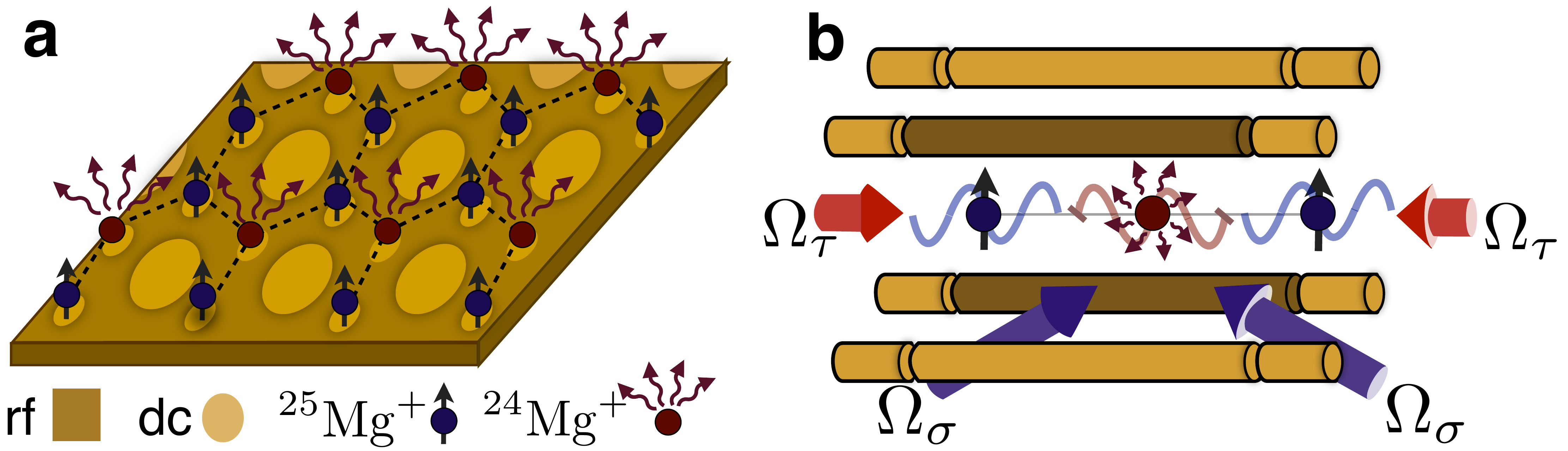}
\caption{ {\bf (a)}  Coulomb crystal in a surface trap. The laser-cooled ions (red) assist the  coherent/dissipative  dynamics of the spins of the physical ions (blue). Both isotopes may be stored in the same individual minima without affecting the resultant lattice geometry. {\bf (b)} Proof-of-principle experiment with three ions in a conventional rf-trap. The red arrows correspond to a standing wave providing the Doppler cooling of the central ion, whereas the blue arrows lead to a running wave tuned to the axial red sideband of the outer ions.   }
\label{fig_scheme}
\end{figure}

{\it The model.--} We consider an array of two types of ions $\{\sigma,\tau\}$ confined in a radio-frequency (rf) trap (Fig.~\ref{fig_scheme}{\bf (a)}). Two hyperfine ground-states  $\{\ket{{\uparrow}},\ket{{\downarrow}}\}$ of the $\sigma$-ions provide the playground for QC/QS, whereas the $\tau$-ions act as an auxiliary gadget to sympathetically cool the crystal. In particular, the $\tau$-ions are Doppler cooled by using a standing wave red-detuned from a dipole-allowed transition with decay rate $\Gamma_{\tau}$~\cite{comment_sw_cooling}, while the $\sigma$-ions are subjected to a spin-phonon coupling obtained from a two Raman beam in a traveling-wave configuration~\cite{wineland_review}. When the laser cooling is strong, the atomic degrees of freedom of the $\tau$-ions can be traced out~\cite{sup_mat}, and one  obtains a master equation for the reduced dynamics of the $\sigma$-spins and the collective vibrations
 \begin{equation}
\label{laser_cooling}
\dot{\mu}=-\ii[H_{\sigma}+H_{\rm ph}+V_{\sigma}^{{\rm ph}},\mu]+\mathcal{\tilde{D}}(\mu),\hspace{2ex}\mu={\rm Tr}_{\tau,{\rm atomic}}\{\rho\}.
\end{equation}
Here,  we have introduced the bare spin and phonon Hamiltonians $H_{\sigma}=\frac{1}{2}\sum_{i}\omega_{0}^{\!\sigma}\sigma_i^z$,  $H_{\rm ph}=\sum_{n}\omega_{n}b_n^{\dagger}b_n,$ where $\omega_{0}^{\!\sigma}$ and $\omega_n$ are the electronic and longitudinal normal-mode frequencies~\cite{comment}. Additionally,  $\sigma_i^z=\ket{{\uparrow_i}}\bra{{\uparrow_i}}-\ket{{\downarrow_i}}\bra{{\downarrow_i}}$,  and   $b_n^{\dagger},b_n^{\phantom{\dagger}}$ are the  operators that create-annihilate phonons. The two crucial ingredients in~\eqref{laser_cooling} for our dissipation-assisted protocol are: 

{\it (i)} A {\it spin-phonon coupling}, provided by the Raman beams tuned to the so-called red sideband~\cite{sup_mat}, leads to
\begin{equation}
\nonumber
V_{\sigma}^{{\rm ph}}\!=\!\sum_{i,n}\mathcal{F}^{\sigma}_{in}\sigma_i^+b_n\ee^{-\ii\omega_{\sigma}t}+\text{H.c.},\hspace{1ex} \mathcal{F}^{\sigma}_{in}=\textstyle{\frac{\ii\Omega_{{\sigma}}}{2}\eta^{\sigma}_n\mathcal{M}_{in}\ee^{\ii\phi_{i}}},
\end{equation}
where $\sigma_i^+=\ket{{\uparrow_i}}\bra{{\downarrow_i}}$, and the sum  is extended to all $\sigma$-ions and normal modes. Here, $\Omega_{\sigma}$ is the Rabi frequency of the Raman beams,  $\omega_{\sigma} ({\bf k}_{\sigma})$ is its  frequency (wavevector), and $\phi_i={\bf k}_{\sigma}\cdot{{\bf r}_i^{\sigma}}$ is defined in terms of the ion position ${{\bf r}_i^{\sigma}}$. The Lamb-Dicke parameter $\eta^{\sigma}_n={\bf k}_{\sigma}\cdot{\bf e}_{\rm d}/\sqrt{2m_{\sigma}\omega_n}$   describes the laser coupling to the $n$-th normal mode, where the $i$-th ion displacement along the  direction ${\bf e}_{\rm d}$ is given by $\mathcal{M}_{in}$, and $m_{\sigma}$ is the ion mass.

{\it (ii)} An {\it effective phonon damping}, provided by the sympathetic Doppler cooling~\cite{sup_mat}, which can be described by
\begin{equation}
\nonumber
\mathcal{\tilde{D}}(\mu)\!=\!\sum_{n}\!\big\{\Gamma_{n}^+\!(b_n^{\dagger}\mu b_n^{\phantom{\dagger}}\!-b_n^{\phantom{\dagger}}b_n^{\dagger}\mu)\!+\!\Gamma_{n}^-\!(b_n^{\phantom{\dagger}}\mu b_n^{\dagger}\!-b_n^{\dagger}b_n^{\phantom{\dagger}}\mu)\!\big\}\!+\!\text{H.c.}
\end{equation}
Here, we have introduced  Lorentzian-shaped cooling/heating couplings, which allow for an experimental control fo the damping of the vibrational modes, and have the expression  $\Gamma_{n}^{\mp}=\sum_l(\half\Omega_{{\tau}}\eta^{\tau}_{n}\mathcal{M}_{ln})^2/(\half\Gamma_{\tau}+\ii(-\Delta_{\tau}\pm\omega_n))$, where we sum over all the $\tau$-ions. In these expressions, we have introduced the laser Rabi frequency $\Omega_{\tau}$, its detuning $\Delta_{\tau}$, and its wavevector ${\bf k}_{\tau}$ that determines  $\eta^{\tau}_n={\bf k}_{\tau}\cdot{\bf e}_{\rm p}/\sqrt{2m_{\tau}\omega_n}$.

The  master equation~\eqref{laser_cooling} describes an array of spins coupled to a set of damped vibrational modes. The  idea now is to use the quanta of these modes, i.e. the phonons, as mediators of a coherent spin-spin interaction. However, in addition to the coherent dynamics, the phonons also provide an indirect coupling to the electromagnetic reservoir  leading to some collective dissipation on the spins. Our goal is to find suitable regimes where these collective effects still allow for QC/QS. To guide this search, note that the two-qubit gates implemented in different laboratories~\cite{gates_review} use nearly-resonant spin-dependent forces, and rely on the motional coherence to suppress the residual spin-phonon entanglement. Since the motional coherence is absent in our case, we must work in the far off-resonant regime~\cite{ms,porras_qs}, where $|\mathcal{F}_{in}^{\sigma}|\ll|\delta_n|\ll\omega_n$, such that $\delta_n=\omega_{\sigma}-(\omega_0^{\sigma}-\omega_n)$. In this regime, motional excitations by the spin-phonon coupling are negligible. We identify below the additional conditions  to tailor the  coherent/dissipative phonon-mediated processes in  presence of laser cooling.

{\it Collective Liouvillian.--} For the  values considered below, the laser-cooling rates  reach  $W_n\approx 10^{-2}\omega_n$. In this case, the cooling is very strong, and the vibrations reach the steady state very fast. Hence, we can apply the theory of Schrieffer-Wolff (SW) transformations for open systems~\cite{open_sw} to trace out the phonons from Eq.~\eqref{laser_cooling}, and obtain an effective Liouvillian
\begin{equation}
\label{eff_me}
\dot{\mu}_{\sigma}=\mathcal{L}_{\rm eff}(\mu_{\sigma})=-\ii[H_{\rm eff},\mu_{\sigma}]+\mathcal{D}_{\rm eff}(\mu_{\sigma}),
\end{equation}
where $\mu_{\sigma}={\rm Tr}_{\rm ph}\{\mu\}$. Here, the coherent  Hamiltonian is 
\begin{equation}
\nonumber
H_{\rm eff}=\sum_{i>j}\big(J_{ij}^{\rm eff}\sigma_i^+\sigma_j^-+\text{H.c.}\big)+\sum_{in}\half B_{in}^{\rm eff}\sigma_i^z,
\end{equation}
which contains the phonon-mediated interactions  of strength $J_{ij}^{\rm eff}$,  which describe processes where a phonon is virtually created, and then absorbed elsewhere in the chain. These interactions can be used to implement gates for QC, or to explore spin models for QS. Additionally, we also find an effective ac-Stark shift, which can be interpreted as an effective magnetic field $B_{in}^{\rm eff}$, arising from the processes where the phonon is created and absorbed by the same ion. Note that the same virtual phonon exchange also introduces an indirect dissipation in~\eqref{eff_me}
\begin{equation}
\nonumber
\begin{split}
\mathcal{D}_{\rm eff}(\mu_{\sigma})=\sum_{i,j}{\Gamma^{\prime}}_{ij}^{\rm eff}&(\sigma_i^+\mu_{\sigma}\sigma_j^--\sigma_j^-\sigma_i^+\mu_{\sigma}+\text{H.c.})\\
+\sum_{i,j}(\Gamma_{ij}^{\rm eff}+{\Gamma'}_{ij}^{\rm eff})&(\sigma_i^-\mu_{\sigma}\sigma_j^+-\sigma_j^+\sigma_i^-\mu_{\sigma}+\text{H.c.}),
\end{split}
\end{equation}
where $\Gamma^{\rm eff}_{ij},\Gamma'^{\rm eff}_{ij}$ are the strengths of the collective processes of spontaneous and stimulated dissipation, respectively. 

 To find the correct regime for QC/QS purposes, we must compare the  time-scales derived from the expressions
\begin{equation}
\nonumber
\label{parameters}
\begin{split}
J_{ij}^{\rm eff}&\!=\!-\!\sum_n\!\frac{\mathcal{F}_{in}^{\sigma}(\mathcal{F}_{jn}^{\sigma})^*}{\tilde{\delta}_n^2+W_n^2}\tilde{\delta}_n,\hspace{2.5ex} B_{in}^{\rm eff}\!=\!-\!\frac{\mathcal{F}_{in}^{\sigma}(\mathcal{F}_{in}^{\sigma})^*}{\tilde{\delta}_n^2+W_n^2}\tilde{\delta}_n(2\bar{n}_n+1),\\
\Gamma_{ij}^{\rm eff}&\!=\!+\!\sum_n\!\frac{\mathcal{F}_{in}^{\sigma}(\mathcal{F}_{jn}^{\sigma})^*}{\tilde{\delta}_n^2+W_n^2}W_n,\hspace{2.ex}{\Gamma'}_{ij}^{\rm eff}\!=\!\sum_n\!\frac{\mathcal{F}_{in}^{\sigma}(\mathcal{F}_{jn}^{\sigma})^*}{\tilde{\delta}_n^2+W_n^2}W_n\bar{n}_n.
\end{split}
\end{equation}
Here, the laser cooling leads to the effective cooling rates $
W_n={\rm Re}\{\Gamma^{-}_n-\Gamma^{+}_n\}$ that damp the ion vibrations, and to a Lamb-type shift of the vibrational frequencies leading to $\tilde{\delta}_n=\delta_n+{\rm Im}\{\Gamma_n^+-(\Gamma_n^{-})^*\}$. Additionally $\bar{n}_n={\rm Re}(\Gamma^{+}_n)/W_n$ are the mean phonon numbers in the steady state of the laser cooling. From these expressions, it is clear that by tuning the ratio $
\mathcal{R}_n=W_n (\bar{n}_n+1)/|\tilde{\delta}_n|$, we  control if the spin interactions prevail over the  dissipation  $\mathcal{R}_n\ll1$, or vice versa $\mathcal{R}_n\gg1$.

{\it Coherent and dissipative generation of entanglement.--} We consider the simplest scenario  to test  our scheme: a three-ion chain in a linear Paul trap (Fig.~\ref{fig_scheme}{\bf (b)}). To use realistic  parameters, we  consider   $^{25}$Mg$^{+}$-$^{24}$Mg$^{+}$-$^{25}$Mg$^{+}$, and set the axial trap frequency to $\omega_z/2\pi=4.1$\hspace{0.2ex}MHz, which is possible by optimizing the trap voltages. The  dipole-allowed transition $\ket{g}=\ket{3S_{1/2}}\leftrightarrow\ket{e}=\ket{3P_{1/2}}$ of $^{24}$Mg$^{+}$, which is characterized by $\lambda_{\tau}\approx 280.3$nm and $\Gamma_{\tau}/2\pi\approx 41.4$\hspace{0.2ex}MHz, shall be used for continuous sympathetic cooling. By applying an external magnetic field, we  can encode the spins in a couple of Zeeman sub-levels  $\ket{F,M}$ of the ground-state manifold of $^{25}$Mg$^{+}$,  e.g. $\ket{{\uparrow}}=\ket{2,2}$ and $\ket{{\downarrow}}=\ket{3,3}$. This leads to a resonance frequency of $\omega^{\sigma}_0/2\pi\approx 1.79$\hspace{0.2ex}GHz, and a negligible decay rate of $\Gamma_{\sigma}/2\pi\approx 10^{-14}$Hz. Finally, a pair of off-resonant lasers drive the axial red-sideband   through an excited state in the $3P_{3/2}$ manifold of $^{25}$Mg$^{+}$, such that $\eta_1^{\sigma}\approx 0.16$.

The isotopic mass ratio  $m_{\tau}/m_{\sigma}\approx 0.96$ implies that the axial vibrational modes are almost unchanged with respect to the homogeneous chain, $\omega_n/2\pi\approx\{4.1,7.1,10.1\}$\hspace{0.2ex}MHz.  To attain a wide range of values for the ratio $\mathcal{R}_n$, we tune the Raman lasers closer to the highest-frequency mode, the so-called egyptian mode, such that the detunings are $\delta_n/2\pi\in\{6.2,3.2,0.3\}$\hspace{0.2ex}MHz. Note that due to the large detuning from the remaining modes, the collective effects will be mediated by the egyptian mode. To sympathetically cool it, we  place the cooling isotope at the middle of the chain, such that it coincides with the node of a  standing-wave laser~\cite{comment_sw_cooling}. This laser has frequency that is red-detuned from the transition, and we set the detuning to be $\Delta_{\tau}=-\Gamma_{\tau}/2$. This leads to a  steady-state mean phonon number $\bar{n}_3=0.65$  independent of the standing-wave Rabi frequency. Therefore, we can modify the Rabi frequency $0.2\leq\Omega_{\tau}/\Gamma_{\tau}\leq 2$ in order to control the cooling rate $W_3$, and thus the ratio $\mathcal{R}_3$, thus exploring regimes of either dominant  dissipation or interactions. We remark that the laser used for  cooling $^{24}$Mg$^{+}$ will be highly detuned from the cooling transition of $^{25}$Mg$^{+}$ (i.e. $\Delta/2\pi\approx 2.7$\hspace{0.2ex}THz), such that the induced decay rate for the considered regime fulfills $\Gamma_{\tau}(\Omega_{\tau}/\Delta)^2/(2\pi)\leq10^{-2}$Hz. Therefore, this laser only contributes with  off-resonant ac-Stark shifts that shall be considered later on. We now explore two possible applications.

\begin{figure}

\centering
\includegraphics[width=0.9\columnwidth]{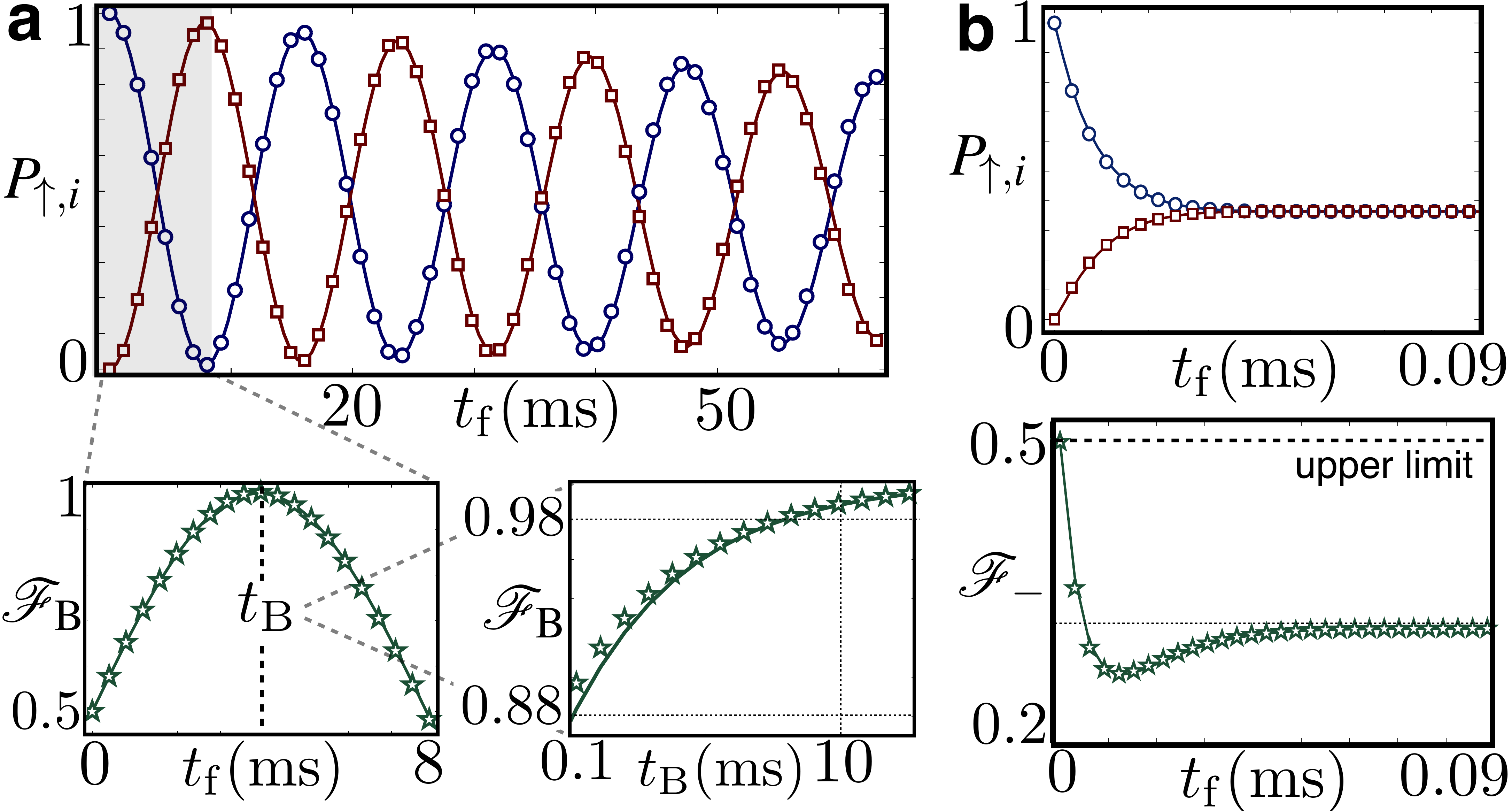}
\caption{ {\bf (a)} In the main panel, we show the coherent flip-flop dynamics for the three-ion setup, when the red sideband is tuned close to the highest-frequency vibrational mode. The solid lines represent the spin populations ($P_{\uparrow,1}$: blue, $P_{\uparrow,3}$: red) given by the original Liouvillian~\eqref{laser_cooling}, while the symbols ($P_{\uparrow,1}$: circles, $P_{\uparrow,3}$: squares) correspond to the effective description~\eqref{eff_me}. In the left lower panel, the fidelity  $\mathcal{F}_{\rm B}=|\langle\psi_{\rm B}|\mu_{\sigma}|\psi_{\rm B}\rangle|$ with the Bell state $\ket{\psi_{\rm B}}=\frac{1}{\sqrt{2}}(\ket{{\uparrow_1\downarrow_3}}-\ii\ket{{\downarrow_1\uparrow_3}})$, for a single flip-flop exchange is displayed. An optimization of the fidelity for different gate times $t_{\rm B}$ is shown in the right lower panel. In both cases, green solid lines represent the complete Liouvillian~\eqref{laser_cooling}, whereas the stars follow from the effective description~\eqref{eff_me}.  {\bf (b)} In the upper panel, the dissipative dynamics under equation~\eqref{laser_cooling} ($P_{\uparrow,1}$: blue solid line, $P_{\uparrow,3}$: red solid line)  and equation~\eqref{eff_me} ($P_{\uparrow,1}$: circles, $P_{\uparrow,3}$: squares) is shown. In the lower panel, we display the fidelity $\mathcal{F}_-=|\langle\phi_{-}|\mu_{\sigma}|\phi_{-}\rangle|$ with the Bell state $\ket{\phi_{-}}=\frac{1}{\sqrt{2}}(\ket{{\uparrow_1\downarrow_3}}-\ket{{\downarrow_1\uparrow_3}})$.  Again, we use green solid lines for the complete Liouvillian~\eqref{laser_cooling}, and  stars for the effective description~\eqref{eff_me}.}
\label{fig_ent}
\end{figure}

{ (a)} Coherent generation of entanglement: Our goal is to use the coherent phonon-mediated interaction in Eq.~\eqref{eff_me} to generate entanglement between the $^{25}$Mg$^{+}$ ions. By setting $\Omega_{\tau}=0.15\Gamma_{\tau}$, we obtain a cooling  rate of  $W_3/\omega_3\approx 4.3\cdot10^{-3}$, such that $\mathcal{R}_3\approx 7\cdot 10^{-3}$, and the Hamiltonian part of the Liouvillian~\eqref{eff_me} thus dominates.  Initializing the spin state in $\ket{\psi_{\sigma}(0)}=\ket{{\uparrow_1\downarrow_3}}$, and  setting $\Omega_{\sigma}\eta_3^{\sigma}\approx 10W_{3}$, such that the distance between the ions is an integer multiple of the effective Raman wavelength,  we obtain the Bell state $\ket{\psi_{\rm B}}=\frac{1}{\sqrt{2}}(\ket{{\uparrow_1\downarrow_3}}-\ii\ket{{\downarrow_1\uparrow_3}})$ 
for $t_{\rm f}\approx 4$\hspace{0.2ex}ms (Fig.~\ref{fig_ent}{\bf (a)}). In the numerical simulations, we have considered a  realistic heating rate  for macroscopic rf-traps of $\Gamma_{\rm ah}\approx 0.1$phonon/ms, by substituting $\Gamma_{n}^+\to\Gamma_{n}^++\Gamma_{\rm an}$ in the dissipator of~\eqref{laser_cooling}.  From this figure, we observe that, even if the process is slower than the usual  gates~\cite{gates_review}, it prevails over the phonon-mediated decoherence leading to errors as low as $\epsilon_{\rm B}\sim 10^{-2}$. Note that such errors are not sufficient for fault-tolerance QC, which require $\epsilon_{\rm ft}\sim 10^{-2}$-$10^{-4}$. On the one hand, we can achieve lower error rates by working with larger detunings. On the other hand, this leads to slower gates, which require an additional scheme to decouple from other sources of decoherence that shall be introduced below. Finally, for the anomalous heating rates in micro-fabricated surface traps $\Gamma_{\rm ah}\approx 1$phonon/ms, the same parameters lead to errors $\epsilon_{\rm B}\approx 2\cdot 10^{-2}$ for $t_{\rm f}\approx 5$\hspace{0.2ex}ms, which illustrates the robustness of our scheme with respect to motional heating. Let us also advance that our protocol might be scaled directly to many ions  for QS~\cite{sup_mat}, which do not require such small error rates.

{ (b)} Dissipative generation of entanglement: A different possibility would be to exploit the collective dissipation in Eq.~\eqref{eff_me} to generate entanglement in the steady state. The idea is to set $\Omega_{\tau}=2\Gamma_{\tau}$, such that   the dissipative part of the Liouvillian~\eqref{eff_me} becomes dominating $\mathcal{R}_3\approx 2.3$, and  we can exploit a superradiant/subradiant phenomenon~\cite{superradiance}. By controlling the ion-ion distance  with respect to the Raman wavelength such that ${\bf k}_{\sigma}\cdot ({\bf r}_1^{\sigma}-{\bf r}_3^{\sigma})=p\pi$, where $p\in\mathbb{Z}$, the decay rates fulfill $\Gamma_{11}^{\rm eff}=\Gamma_{33}^{\rm eff}=\Gamma_{\rm eff}$, and $\Gamma_{13}^{\rm eff}=\Gamma_{31}^{\rm eff}=(-1)^p\Gamma_{\rm eff}$ (equally for ${\Gamma^{\prime}}_{ij}^{\rm eff}$). In this limit, the dissipator in~\eqref{eff_me} can be written as 
\begin{equation}
\nonumber
\mathcal{D}_{\rm eff}(\mu_{\sigma})=L_-\mu_{\sigma}L_-^{\dagger}+L_+\mu_{\sigma}L_+^{\dagger}-L_-^{\dagger}L_-\mu_{\sigma}-L_+^{\dagger}L_+\mu_{\sigma}+\text{H.c.},
\end{equation}
where we have introduced the collective jump operators $L_-=\sqrt{\Gamma_{\rm eff}(\bar{n}_3+1)}(\sigma_1^-+(-1)^p\sigma_3^-)$, and $L_+=\sqrt{\Gamma_{\rm eff}\bar{n}_3}(\sigma_1^++(-1)^p\sigma_3^+)$. One can check that  the symmetric/antisymmetric Bell states $\ket{\phi_{\pm}}=\frac{1}{\sqrt{2}}(\ket{{\uparrow_1\downarrow_3}}\pm\ket{{\downarrow_1\uparrow_3}})$ are dark states of these jump operators for $p$ odd, or $p$ even, respectively. These are the so-called sub-radiant decay channels~\cite{superradiance_review}, which allows us to get a mixed stationary state that is partially entangled.  In particular, starting 
 from $\ket{\psi_{\sigma}(0)}=\ket{{\uparrow_1\downarrow_3}}$ for $p$ even, and   $\Omega_{\sigma}\eta_3^{\sigma}\approx W_{3}$, we obtain a decoherence-free entangled steady-state  $\ket{\phi_-}$ for $t\gg t_{\rm ss}\approx50$\hspace{0.2ex}$\mu$s with fidelities around 30$\%$ (Fig.~\ref{fig_ent}{\bf (b)})\cite{comment_entanglement} . Note that this {\it phononic subradiance} is not affected by limitations in the ratio of the ion-ion distance with respect to the wavelength of the emitted light. The collective nature of the vibrations that mediate the subradiance allows to surpass the limitations of the pioneering trapped-ion experiments~\cite{sup_ions}. We also note that the ultimate limit of $50\%$ cannot be achieved due to the thermal contribution to Eq.~\eqref{eff_me}. However,  schemes originally formulated for cavities~\cite{steady_entanglement} can  be adapted for our trapped-ion setting to reach unit fidelities. 

 Let us emphasize that, although we have considered a particular example, the scheme is also applicable  to other ion species. For the regime of dominant dissipation, any ion  will work equally well. Conversely, for dominant coherent interactions, the required strong sympathetic-cooling strengths and detunings are likely to be optimized for crystals with two light isotopes. At this point, it is also worth commenting that the strong rates provided by standing-wave laser cooling are required to obtain the target  states in time-scales which are not prohibitively large. In light of the results shown in Fig.~\ref{fig_ent}, the regime of coherent interactions necessarily requires standing-wave cooling. Conversely, the regime of leading dissipation is faster, and may also work with the more standard traveling-wave cooling.  Let us note, however, that  the experiments~\cite{slowly_moving_standing_wave} show that standing-wave cooling with a precise  positioning of the ions with respect to the standing wave is possible.
 
{\it Sources of noise.--}  In addition to the motional heating,  other sources of noise  become relevant for the  time-scales of  the above protocols  $0.1$-$10$\hspace{0.2ex}ms. In fact, fluctuating magnetic fields and laser intensities, together with thermal noise, lead to the dephasing term $H_{\rm n}=\sum_i\frac{1}{2}\big(\sum_nB^{\rm eff}_{in}+F_i(t)\big)\sigma_i^z$.
Here, $B^{\rm eff}_{in}$ in Eq.~\eqref{eff_me} introduces noise via  fluctuations over the phonon steady state, and $F_i(t)$ is a random process that models the noise of external magnetic fields, or uncompensated ac-Stark shifts. This random noise, which typically has a short correlation time $\tau_{\rm c}$, leads to a dephasing rate $\Gamma_{\rm d}/2\pi\sim$0.1-1$\,$kHz~\cite{sup_mat}. 

For any practical implementation of the proposed protocol, this dephasing should be carefully considered. The standard approach for prolonging the coherence  of a system consists of a sequence of refocusing pulses, a technique known as pulsed dynamical decoupling~\cite{pdd}. Another approach, known as continuous dynamical decoupling, produces a similar effect by  continuous drivings~\cite{cdd_trapped_ions, cdd_ions}. As recently demonstrated experimentally~\cite{cdd_ions_exp}, this techniques allows to implement robust 2-qubit gates exploiting a single sideband~\cite{cdd_ions}. As a byproduct, we show that this tool
 allows for slower gates, and thus smaller errors in principle. Moreover, it also provides a new gadget to tailor the collective Liouvillian~\eqref {eff_me}.
 
We apply a  continuous driving resonant with the spins, such that the bare spin Hamiltonian reads
$
H_{\sigma}=\half\sum_{i}\omega^{\sigma}_0\sigma_i^z+\left(\Omega_{\rm d}\sigma_i^+\cos(\omega_{\rm d}t)+\text{H.c.}\right)$ with $\omega_{\rm d}=\omega^{\sigma}_0$. In this regime,  a modified SW transformation leads to  $\dot{\mu}_{\sigma}=\tilde{\mathcal{L}}_{\rm eff}(\mu_{\sigma})$, where
\begin{equation}
\label{eff_me_ising}
\tilde{\mathcal{L}}_{\rm eff}(\mu_{\sigma})=-\ii[\tilde{H}_{\rm eff}+\tilde{H}_{\rm n},\mu_{\sigma}]+\tilde{\mathcal{D}}_{\rm eff}(\mu_{\sigma})+\tilde{\mathcal{D}}_{\rm n}(\mu_{\sigma}).
\end{equation}
In the limit of a strong driving~\cite{sup_mat}, the  Hamiltonian above corresponds to an interacting  Ising model 
\begin{equation}
\nonumber
\tilde{H}_{\rm eff}=\sum_{i>j}\tilde{J}_{ij}^{\rm eff}\sigma_i^x\sigma_j^x+\sum_{in}\half \Omega_{\rm d}\sigma_i^x,
\end{equation}
and we obtain a collective phonon-mediated  dephasing  
\begin{equation}
\nonumber
\tilde{\mathcal{D}}_{\rm eff}(\mu_{\sigma})=\sum_{i,j}\tilde{\Gamma}_{ij}^{\rm eff}(2\sigma_i^x\mu_{\sigma}\sigma_j^x-\sigma_j^x\sigma_i^x\mu_{\sigma}-\mu_{\sigma}\sigma_j^x\sigma_i^x).
\end{equation}
Assuming that the surface-trap array is designed such that ${\bf k}_{\rm L}\cdot{\bf r}_i^{\sigma}=2\pi p$ with $p\in\mathbb{Z}$, we emphasize that the  interaction  strengths and dissipation rates,
\begin{equation}
\nonumber
\tilde{J}_{ij}^{\rm eff}\!=\!-\!\sum_n\!\frac{|\mathcal{F}_{in}^{\sigma}\mathcal{F}_{jn}^{\sigma}|}{\tilde{\delta}_n^2+W_n^2}\frac{\tilde{\delta}_n}{2}, \hspace{2ex}
\tilde{\Gamma}_{ij}^{\rm eff}\!=\!\sum_n\!\frac{|\mathcal{F}_{in}^{\sigma}\mathcal{F}_{jn}^{\sigma}|}{\tilde{\delta}_n^2+W_n^2}\frac{W_n}{2}(\bar{n}_n+\half),
\end{equation}
lead to a very similar control parameter $
\tilde{\mathcal{R}}_n=W_n (\bar{n}_n+\half)/|\tilde{\delta}_n|$. Accordingly, under the same assumptions  as we considered above, we can interpolate between regimes where the coherent Ising interactions dominate over the collective dephasing, or vice versa.
In addition to the new range of possibilities offered by this collective Liouvillian, note that the dephasing noise terms contribute to the new Liouvillian~\eqref{eff_me_ising} with
\begin{equation}
\nonumber
\tilde{H}_{\rm n}=\sum_i\frac{1}{2}\tilde{\Omega}_{\rm d}\sigma_i^x, \hspace{2ex}\tilde{\mathcal{D}}_{\rm n}(\mu_{\sigma})=\sum_{i}\sum_{\alpha=y,z}\half\tilde{\Gamma}_{\rm d}\left(\sigma_i^{\alpha}\mu_{\sigma}\sigma_i^{\alpha}-\mu_{\sigma}\right),
\end{equation} 
where we have assumed that the noise is local, and introduced $\tilde{\Omega}_{\rm d}=({B}_{jn}^{\rm eff}\tilde{\delta}_n\tau_{\rm c}+\Gamma_{\rm d})/2\Omega_{\rm d}\tau_{\rm c}$, and $\tilde{\Gamma}_{\rm d}=\Gamma_{\rm d}/(\Omega_{\rm d}\tau_{\rm c})^2$ in the limit of a strong driving $\Omega_{\rm d}\tau_{\rm c}\gg 1$~\cite{sup_mat}.
According to the above constraints,  these noisy terms are  suppressed by a sufficiently-strong driving. To be more specific, for noise correlation times on the order of $\tau_{\rm c}=10^{-2}/\Gamma_{\rm d}$ and  detunings $\delta_n/2\pi\sim$0.1-1\hspace{0.2ex}MHz, it suffices to apply drivings with $\Omega_{\rm d}/2\pi\approx 10$\hspace{0.2ex}MHz to reduce the  noise by more than two orders of magnitude. Therefore, we can decouple from the noise  efficiently, while preserving the collective part of the Liouvillian for QC/QS. In contrast to Fig.~\ref{fig_ent}{\bf (a)}, the new Liouvillian~\eqref{eff_me_ising} allows for the coherent generation of all four Bell states.

{\it Conclusions.--} We have proposed a  scheme based on sympathetic cooling to overcome the anomalous heating in surface traps, while allowing for QC/QS. The sympathetic cooling becomes a tool to tailor the collective effects of the Liouvillian. By controlling  a single parameter, namely the laser-cooling power, we have shown how the Liouvillian interpolates between regimes of dominating  coherent interactions/collective dissipation, both of which allow for generation of entanglement. Moreover, we have  this control  can also  be exploited for coherent/dissipative many-body QS.

{\it Acknowledgements.--}  This work was supported by  PICC, DFG (SCHA973/1-6), and  by the Alexander von Humboldt Foundation. A.B. thanks FIS2009-10061, and QUITEMAD. We thank A. Albrecht and J. Almeida for useful discussions.


\newpage
\vspace*{150ex}

\hypertarget{sm}{\section{Supplemental Material: \\ {\it Dissipation-Assisted Quantum Information Processing with Trapped Ions }}}

{\it Dissipative ion crystals.--} We consider a collection of atomic ions confined in radio-frequency traps~\cite{wineland_review_sm}. In particular,  we will explore dissipation-assisted protocols for an  array of ${N}_{\rm t}=N_{\sigma}+N_{\tau}$ trapped ions in equilibrium positions $\{{\bf r}_i^{\sigma},{\bf r}_l^{\tau}\}$. The geometry of the array will depend on the particular trap under consideration. For micro-fabricated surface traps, we will consider arbitrary geometries (Fig.~\ref{fig_scheme}{\bf (a)}), whereas for the more standard rf-traps, we restrict to one-dimensional  chains (Fig.~\ref{fig_scheme}{\bf (b)}). A fraction of the ions, $N_{\tau}$, is laser cooled  via a dipole-allowed transition $\ket{g}\leftrightarrow\ket{ e}$ with frequency ${\omega}_0^{\tau}$ and decay rate $\Gamma_{\tau}$, providing the sympathetic cooling of the remaining ions, $N_{\sigma}$. Two hyperfine ground-states  of the $\sigma$-ions $\{\ket{{\uparrow}},\ket{{\downarrow}}\}$ form the spins/qubits for QS/QC purposes,  such that their decay rate  is negligible, and their energy spacing is $\omega_0^{\sigma}$ ($\hbar=1$).  To minimize the action of the cooling laser   on the spins,   one may use beams  tightly focused on the ${\tau}$-species, or exploit two different ion species/isotopes.

As customary in the theory of open quantum-optical systems~\cite{breuer_book},  one obtains the master equation   
\begin{equation}
\dot{\rho}=-\ii[H,\rho]+\mathcal{D}(\rho),
\end{equation} after tracing out the  electromagnetic bath. Here, we have introduced the Hamiltonian $H$ and  dissipative $\mathcal{D}(\rho)$ parts. Let us start by describing the Hamiltonian 
\begin{equation}
H=H_{\tau}+H_{\sigma}+H_{\rm ph}+V_{\sigma}^{{\rm ph}}+V_{\tau}^{{\rm ph}}.
\end{equation}
Here, $H_{\tau}=\frac{1}{2}\sum_l\omega^{\!{\tau}}_{0}\tau_{l}^z$, $H_{\sigma}=\frac{1}{2}\sum_{i}\omega_{0}^{\!\sigma}\sigma_i^z$, and $H_{\rm ph}=\sum_{n}\omega_{n}b_n^{\dagger}b_n,$
represent the atomic degrees of freedom of the laser-cooled $\tau$-ions, the (pseudo)spins of the $\sigma$-ions, and the vibrational excitations of the ion crystal, respectively. We have introduced  $\tau_l^z=\ket{{e}_l}\bra{e_l}-\ket{g_l}\bra{g_l}$, $\sigma_i^z=\ket{{\uparrow_i}}\bra{{\uparrow_i}}-\ket{{\downarrow_i}}\bra{{\downarrow_i}}$,  and the  creation/annihilation  operators  $b_n^{\dagger},b_n^{\phantom{\dagger}}$ for a  particular phonon branch with frequencies $\omega_n$. Additionally, we include a spin-phonon coupling that is provided by  a stimulated Raman transition~\cite{wineland_review_sm} tuned to the so-called vibrational red-sideband
\begin{equation}
\label{resolved_sideband_sm}
V_{\sigma}^{{\rm ph}}\!=\!\sum_{in}\mathcal{F}^{\sigma}_{in}\sigma_i^+b_n\ee^{-\ii\omega_{\sigma}t}+\text{H.c.},\hspace{1ex} \mathcal{F}^{\sigma}_{in}=\textstyle{\frac{\ii\Omega_{{\sigma}}}{2}\eta^{\sigma}_n\mathcal{M}_{in}\ee^{\ii\phi_{i}}},
\end{equation}
where we re-write the definitions already used in the main text for convenience. Here, $\sigma_i^+=\ket{{\uparrow_i}}\bra{{\downarrow_i}}$. Here, $\Omega_{\sigma}$ is the two-photon Rabi frequency,  $\omega_{\sigma} ({\bf k}_{\sigma})$ is the Raman frequency (wavevector), and $\phi_i={\bf k}_{\sigma}\cdot{{\bf r}_i^{\sigma}}$. The Lamb-Dicke parameter $\eta^{\sigma}_n={\bf k}_{\sigma}\cdot{\bf e}_{\rm d}/\sqrt{2m_{\sigma}\omega_n}$   describes the laser coupling to the $n$-th  mode with displacements $\mathcal{M}_{in}$ along ${\bf e}_{\rm d}$, where $m_{\sigma}$ is the ion mass.  This spin-phonon coupling arises from the dipole laser-ion interaction after expanding $\eta_{n}^{\sigma}\ll 1$, such that $\omega_{\sigma}\approx \omega_0^{\sigma}-\omega_n$. We set $|\Omega_{\sigma}|\ll\omega_n$ to neglect the contribution of other terms (i.e. carrier and blue sideband)  from the laser-ion interaction.

 We laser cool the $\tau$-species in the node of a standing wave~\cite{laser_cooling_cirac}, which shall be red-detuned with respect to the atomic transition, and can be described by 
\begin{equation}
\label{non_resolved_sideband_sm}
V_{\tau}^{{\rm ph}}=\sum_{ln}\mathcal{F}^{\tau}_{ln}\tau_l^+Q_n\ee^{-\ii\omega_{\tau}t}+\text{H.c.}, \hspace{1ex} \mathcal{F}^{\tau}_{ln}=-\textstyle{\frac{\Omega_{{\tau}}}{2}\eta^{\tau}_n\mathcal{M}_{ln}},
\end{equation}
where  $\tau_l^+=\ket{{e_l}}\bra{{g_l}}=(\tau_l^-)^{\dagger}$, $Q_n=b_n^{\phantom{\dagger}}+b_n^{\dagger}$, and the remaining parameters are defined as below  Eq.~\eqref{resolved_sideband_sm}. Note that the differences between Eqs.~\eqref{resolved_sideband_sm} and~\eqref{non_resolved_sideband_sm} are due to the different regimes $\Gamma_{\sigma}\ll\omega_n\ll\Gamma_{\tau}$, which forbid resolving the sidebands of the dipole-allowed transition of the $\tau$-ions (i.e. Doppler cooling regime). Additionally, the component of the laser-ion interaction that drives the carrier vanishes at the node of the standing wave. This allows us to consider strong Rabi frequencies $\Omega_{\sigma}$ to optimize the cooling rates in the regime of interest.

 Finally, the dissipator including recoil effects~\cite{recoil}, can be described as the sum of two terms
 \begin{equation} 
 \mathcal{D}(\rho)=\mathcal{D}_0(\rho)+\mathcal{D}_1(\rho).
 \end{equation}
 Here, $\mathcal{D}_0(\rho)$ is the usual dissipation super-operator in Lindblad form~\cite{lindblad} for a two-level atom at rest
\begin{equation}
\mathcal{D}_0(\rho)=\sum_l\half\Gamma_{\tau}\left(\tau_l^{-}\rho\tau_l^{+}-\tau_l^{+}\tau_l^{-}\rho\right)+\text{H.c.},
\end{equation}
 where  the typical ion distances forbid collective dissipative effects. Additionally $
\mathcal{D}_1(\rho)$ which describes recoil effects
\begin{equation}
\mathcal{D}_1(\rho)=\sum_{lnm}\half\Gamma_{\tau,nm}\tau_l^{-}\big(Q_n\rho Q_m-Q_nQ_m\rho\big)\tau_l^{+}+\text{H.c.},
\end{equation}
where we have introduced $\Gamma_{\tau,nm}=2\Gamma_{\tau}\alpha_q\eta^{\tau}_n\eta^{\tau}_m\mathcal{M}_{ln}\mathcal{M}_{lm}$, and $\alpha_q=(1+3q^2)/10(1+q^2)$, such that $q=0,\pm1$ depends on the linear/circular polarization of the emitted photon.

In addition to the Doppler-regime condition $\omega_n\ll\Gamma_{\tau}$, we further impose that  $\mathcal{F}^{\tau}_{ln}\ll\omega_n,\Omega_{\tau}$. In this case,  the laser-cooled ions reach the steady state very fast, and can be integrated out~\cite{laser_cooling_cirac}, which leads to the master equation
\begin{equation}
\label{laser_cooling_sm}
\dot{\mu}=-\ii[H_{\sigma}+H_{\rm ph}+V_{\sigma}^{{\rm ph}},\mu]+\mathcal{\tilde{D}}(\mu),\hspace{2ex} \mu={\rm Tr}_{\tau,{\rm at}}\{\rho\}
\end{equation}
which is the starting point of our work in Eq.~\eqref{laser_cooling}. The  effective dissipator describing the laser cooling of the vibrational modes is
\begin{equation}
\mathcal{\tilde{D}}(\mu)\!=\!\sum_{n}\!\big\{\Gamma_{n}^+\!(b_n^{\dagger}\mu b_n^{\phantom{\dagger}}\!-b_n^{\phantom{\dagger}}b_n^{\dagger}\mu)\!+\!\Gamma_{n}^-\!(b_n^{\phantom{\dagger}}\mu b_n^{\dagger}\!-b_n^{\dagger}b_n^{\phantom{\dagger}}\mu)\!\big\}\!+\!\text{H.c.},
\end{equation}
where the effective rates are expressed as $\Gamma^{\pm}_n=S(\mp\omega_n)$, and $S(\omega_n)=\int_0^{\infty} ds\ee^{\ii\omega_ns}\langle F_n(s)F_n(0)\rangle_{\rm ss}$ is the steady-state fluctuation spectrum of $F_n=\sum_l\mathcal{F}_{ln}^{\tau}(\tau^{+}_l+\tau^-_l)$. At this level, we  introduce the heating by  $\Gamma^{+}_n\to\Gamma^{+}_n=S(-\omega_n)+\Gamma^{\rm ah}_n$, where $\Gamma^{\rm ah}_n$ is the anomalous heating rate. The cooling rates and mean phonon numbers are 
$
W_n={\rm Re}\{\Gamma^{-}_n-\Gamma^{+}_n\},\hspace{1ex} \bar{n}_n={\rm Re}(\Gamma^{+}_n)/W_n$. Therefore, it is straightforward to see that we can overcome the  heating  by shaping  the laser-cooling fluctuation spectrum such that $W_n>0$, and obtain an overall cooling.

\vspace{1ex}
{\it Modeling the noisy dynamics.--} Let us now describe in more details how to take into account possible sources of noise, which appear for the time-scales of interest in addition to the anomalous heating. We will consider three possible sources: {\it (i)} The fluctuations around the state-state mean phonon number $\bar{n}_n$ will cause a pure dephasing of a thermal origin. {\it (ii)} Fluctuations of non-shielded Zeeman shifts will induce a pure dephasing of a magnetic origin. {\it (iii)} Fluctuations of non-compensated ac-Stark shifts also induce a pure dephasing, whose major contribution may be cuased by instabilities in the laser intensity of the cooling lasers. These three terms can be modeled by
\begin{equation}
H_n=\sum_i\frac{1}{2}\big(\sum_nB^{\rm eff}_{in}+F_i(t)\big)\sigma_i^z,
\end{equation} 
where $B^{\rm eff}_{in}$ in Eq.~\eqref{eff_me} yields the thermal noise, and $F_i(t)$ is a random process for the magnetic/laser-intensity dephasing. We assume a local Gaussian noise with a short correlation time $\tau_{\rm c}$, which determines the stochastic average of two-time correlators $\langle F_j(s)F_i\rangle_{\rm st}=\frac{2\Gamma_{\rm d}}{\tau_{\rm c}}\ee^{-s/\tau_{\rm c}}\delta_{ji}$, and in turn the dephasing rate of the spin dynamics $\Gamma_{\rm d}=\half\int_0^{\infty}{\rm d}s\langle F_j(s)F_j\rangle_{\rm st}=1/2T_2$. Let us note that typical decoherence times in trapped-ion experiments are $T_2\approx 1$-10\hspace{0.2ex}ms ($\Gamma_{\rm d}\approx 0.05$-$0.5 $\hspace{0.2ex}kHz), and  that  the regimes considered above yield  $B_{in}^{\rm eff}/\bar{n}_n\approx 0.3$-$3$\hspace{0.2ex}kHz. Comparing these values to the time-scales of the dissipation-assisted protocols ($\sim$$0.1$-$10$\hspace{0.2ex}ms), it emphasizes that we need a scheme to actively decouple from this noise.

 A partial solution  would be the use of  states that are insensitive to linear Zeeman shifts,  such as $\ket{{\uparrow}}=\ket{2,1}$ and $\ket{{\downarrow}}=\ket{3,1}$ at a field of $B_0=213\hspace{0.2ex}$G for $^{25}$Mg$^{+}$. However, since we still need to mitigate the other sources of dephasing, we will exploit  a different mechanism. We introduce a continuous driving of the spins, such that  the bare spin Hamiltonian 
 \begin{equation}
H_{\sigma}=\half\sum_{i}\omega^{\sigma}_0\sigma_i^z+\left(\Omega_{\rm d}\sigma_i^+\cos(\omega_{\rm d}t)+\text{H.c.}\right),
\end{equation} 
may be provided by a microwave source.
Here, the  driving parameters are $\omega_{\rm d}=\omega^{\sigma}_0$, and  $|\Omega_{\rm d}|\ll\omega_{\sigma}$.
Since the resonance frequencies are $\omega_{\sigma}/2\pi\approx 1$\hspace{0.2ex}GHz, the driving can still be strong enough to fulfill {\it strong-driving conditions}
\begin{equation}
 \{W_n,\tilde{\delta_n}\}\ll |\Omega_{\rm d}|, \hspace{2ex} \Gamma_{\rm d},\tau_{\rm c}^{-1}\ll |\Omega_{\rm d}|.
\end{equation}
The first of these conditions allows us to use a modified Schriefer-Wolff transformation, which leads us to the new phonon-mediated terms $\tilde{H}_{\rm eff}$ and $\tilde{\mathcal{D}}_{\rm eff}$ of the Liouvillian in Eq.~\eqref{eff_me_ising} of the main text, after neglecting of-resonant contributions for such a strong driving. The second condition allows us to obtain the effect of the residual noise $\tilde{H}_{\rm n}$ and $\tilde{\mathcal{D}}_{\rm n}$ in  Eq.~\eqref{eff_me_ising}, by using a Born-Markov approximation
\begin{equation}
\dot{\hat{\rho}}=-\int_0^{\infty}{\rm d}s\langle[\hat{H}_{\rm n}(t),[\hat{H}_{\rm n}(t-s),\hat{\rho}(t)]]\rangle_{\rm st},
\end{equation}
where we perform an stochastic average, and work in the interaction picture with respect to the resonant diving $\hat{H}_{\rm n}(t)=UH_{\rm n}U^{\dagger}$, where $U={\rm exp}\{\ii t\sum_i\half\Omega_{\rm d}\sigma_i^x\}$.

{\it Many-body physics.--} The collective Liouvillians in Eqs.~\eqref{eff_me} and~\eqref{eff_me_ising} define our toolbox for dissipation-assisted QS. In the main text, we have considered quantum information processing by means of an isolated vibrational mode that mediates the collective dynamics. In the many-ion scenario, this would lead to fully-connected spin networks that become very interesting in the presence of magnetic frustration~\cite{network}. However, to achieve strongly-correlated models,  it is better to work with a full vibrational branch of a small frequency width. The small ratios $\mathcal{R}_n\sim 10^{-3}$ obtained for trapping frequencies $\omega_{\rm t}/2\pi\approx 10$MHz, indicate that the coherent dynamics can  be dominating also in this case, while  minimizing the  heating. This would allow for the QS of exotic  models with 3-body interactions~\cite{exotic} or topological order~\cite{kitaev} in surface ion traps.

These many-body QS  focus on   the Hamiltonian while minimizing the influence of the environment. However,  the dissipative  dynamics may also lead to  interesting many-body phenomena~\cite{dissipative_many_body}. The possibility to control the relative strength of the coherent/dissipative parts in~\eqref{eff_me} and~\eqref{eff_me_ising} is very appealing in this respect. In particular, we note that for intermediate driving strengths $\Omega_{\rm d}$, competing dissipative terms in~\eqref{eff_me_ising} may lead to purely-dissipative quantum phase transitions.


\begin{references}
\bibitem{comment_sw_cooling}
 Standing-wave laser cooling is considered for generality. However, for the protocol of dissipative QIP, the standard travelling-wave cooling may suffice (see a more detailed discussion below).
\bibitem{comment}
 The scheme works for any vibrational branch of the  crystal (e.g. axial and radial in a chain) provided that the Coulomb couplings and sympathetic cooling rates are of comparable magnitude.
\bibitem{comment_entanglement}
We have also computed the an entanglement measure for the steady state, namely the logarithmic negativity~\cite{log_neg} ${LN}(\mu_{\sigma}^{\rm ss})\approx0.12$, to rule out that this corresponds to a product state.
\end{references}
\end{document}